\documentclass[runningheads]{llncs}

\usepackage[T1]{fontenc}
\usepackage[utf8]{inputenc}
\usepackage{amsmath, amssymb, amsfonts}
\usepackage{graphicx}
\graphicspath{{figures/}}
\usepackage{float}
\usepackage{booktabs}
\usepackage{tikz}
\usetikzlibrary{arrows.meta,calc,positioning,shapes.geometric}
\usepackage{hyperref}
\usepackage{xurl}
\usepackage{cleveref}
\usepackage{xcolor}
\usepackage[expansion=false]{microtype}
\usepackage{orcidlink}           


\begin{document}

\title{Agentic AI-Powered Re-Identification:\\
  An Emerging, Scalable Threat to Mobility Microdata Privacy}

\titlerunning{Agentic AI-Powered Re-Identification}

\author{
  Oscar Thees\orcidlink{0009-0001-9378-4988}\inst{1}\thanks{$^*$ Equal contribution.}
  \and
  Roman M\"{u}ller\orcidlink{0009-0007-2142-3896}\inst{1}$^*$
  \and
  Matthias Templ\orcidlink{0000-0002-8638-5276}\inst{1}
}

\authorrunning{O. Thees, R. M\"{u}ller, M. Templ}

\institute{
  University of Applied Sciences and Arts Northwestern Switzerland (FHNW),\\
  Riggenbachstrasse 16, 4600 Olten, Switzerland\\
  \email{oscar.thees@fhnw.ch}
}

\maketitle

\begin{abstract}
The widespread collection of fine-grained location data by commercial data
brokers creates a re-identification risk that is not widely recognised by the public.
While prior research has established that mobility traces are highly unique
and that individuals can, in principle, be identified from a handful of
spatio-temporal points, such attacks have historically required significant
manual effort from skilled analysts, limiting their practical scale.

In this feasibility study, we demonstrate in a real world setting that agentic AI fundamentally changes this threat model. We present an end-to-end pipeline in which large language model agents autonomously search the open web, cross-reference
public records and social media, and resolve raw coordinate sequences to
candidate identities -- without human intervention. We evaluate the pipeline on a spatio-temporal dataset containing simulated location points anchored at and around true home and work addresses, focusing on a high-risk disclosure scenario. 
Our results demonstrate that, from spatio-temporal data and public sources alone, our agentic AI successfully re-identified 18 of the 25 re-identifiable individuals (72\%) and 18 of 43 cases overall (41.9\%).

We discuss implications for Statistical Disclosure Control (SDC) practice and
outline the near-future escalation that data custodians and regulators
must anticipate. De facto anonymity -- an implicit foundation of SDC practice -- is shifting. Agentic AI strengthens the case that re-identification is \emph{reasonably likely} by any means under the GDPR Recital-26 standard, at costs of minutes-and-dollars per target.

\end{abstract}

\section{Introduction}
\label{sec:introduction}

Every day, hundreds of millions of smartphones emit precise GPS coordinates
that are silently collected by a largely invisible industry of commercial
data brokers~\cite{FTCGravy2024,FTCMobilewalla2024,MeineckDachwitz2024}.
These coordinates -- timestamped and collected at sub-minute resolution
-- are packaged into mobility datasets and traded on open markets under
the label of anonymised data. In practice, anonymisation is limited to
the removal of direct identifiers; the fine-grained traces (themselves strong quasi-identifiers) remain at full resolution, often
alongside device-level metadata~\cite{MeineckDachwitz2024}. Recent analyses confirm that this practice -- routinely labelled \emph{anonymisation} but, under GDPR's identifiability standard, more properly pseudonymisation -- remains the industry default despite its well-documented technical inadequacy~\cite{Gadotti2024}. Under the GDPR identifiability standard~\cite{GDPR2016Recital26}, such spatio-temporal data is personal data whenever re-identification is reasonably likely by any means -- a position reinforced by recent CJEU jurisprudence such as the IAB Europe ruling~\cite{CJEUIAB2024}.
The scale is substantial: a single broker dataset documented by
journalists contained more than 50 billion location pings from 12 million
U.S. smartphones over a few months~\cite{ThompsonWarzel2019}.

The privacy risk embedded in such data is well-documented \cite{deMontjoye2013}. 
Standard anonymisation techniques -- suppression, spatial generalisation,
trajectory perturbation -- offer limited protection against a determined
adversary who can cross-reference external information
sources~\cite{Zang2011,Primault2019}. Yet despite this theoretical
clarity, practical large-scale re-identification has remained a labour-intensive task confined to skilled analysts~\cite{Ko2026}. The bottleneck has not been the uniqueness of mobility traces, but the human effort required to link
raw coordinate traces to a real-world identity and validate the result.

\paragraph{The agentic AI inflection point.}
We argue that recent advances in large language model (LLM) agents change this picture
decisively. A modern LLM agent can autonomously decompose a complex
investigative task into sub-goals, issue queries to search engines, retrieve
and parse web pages, reason over the returned evidence, and iterate until a
conclusion is reached -- all without human involvement~\cite{Yao2023ReAct}.
What previously required an experienced open-source intelligence (OSINT)
analyst working for hours can now be delegated to an agent that
completes equivalent steps in minutes and can be parallelised across
many targets at low marginal cost~\cite{Staab2024Beyond,Lermen2026}.

\paragraph{De facto anonymity and the feasibility threshold.}
AI challenges de facto anonymity by expanding the set of identification
methods that are realistically feasible in practice. Because anonymity
depends not on the abstract impossibility of re-identification but on
whether identification is \emph{reasonably likely} using available means~\cite{GDPR2016Recital26},
improvements in AI-driven linkage, inference, and retrieval materially
raise disclosure risk. At the same time, generative AI lowers the
expertise required to perform such attacks, making capabilities once
limited to specialised analysts increasingly available to a broader
range of actors \cite{Li2026Anthropic,Fawzy2026VibeCoding}.

\paragraph{Our contribution.}
In this paper we make the following contributions:

\begin{itemize}
  \item We develop and evaluate an \emph{agentic re-identification pipeline} that takes
        raw coordinate traces as input and resolves them to
        candidate identities using web-based open-source intelligence tools
        (\Cref{subsec:pipeline}).

  \item We show that the agentic pipeline correctly named 18 of the 25 re-identifiable individuals (72\%) and 18 of all 43 cases (41.9\%) at an average cost of \$2.24 and 17 minutes per target. Further, we discuss the effectiveness and scaling capabilities of the pipeline compared to a human analyst
        (\Cref{sec:results}).

  \item We sketch two near-future escalation vectors -- agentic outreach to confirm candidates and collect additional re-identifiable information, and locally-executed open-weight models -- and outline what each implies for SDC if attacker effort continues to fall (\Cref{sec:conclusion}).

\end{itemize}

The ethical study design choices that resolve the tension between scientific rigour and privacy protection (i.e., consent-based recruitment, simulated traces around true home and work anchors, exclusively public sources during attribution, and non-release of both the simulated data and the underlying code and prompts) are detailed in \Cref{subsec:study_design}. \Cref{sec:related_work} reviews related work, \Cref{sec:methodology} present the pipeline and the setup, \Cref{sec:results} reports results, and \Cref{sec:conclusion} discusses implications.

\section{Related Work}
\label{sec:related_work}

\paragraph{Re-identification of mobility data.}
Journalistic investigations have demonstrated that commercially traded location
data can often be linked to individuals, even when the datasets contain
neither direct identifiers nor sociodemographic information~\cite{ThompsonWarzel2019,MeineckDachwitz2024}. This is especially concerning when such data are collected or disseminated without meaningful user consent~\cite{FTCGravy2024,FTCMobilewalla2024}.

The re-identification risk of mobility data has also been extensively studied in 
academia. For example, de Montjoye et~al.~\cite{deMontjoye2013} demonstrated that four
spatio-temporal points suffice to single out $95\%$ of individuals in a
mobile phone dataset of 1.5 million users. Farzanehfar et~al.~\cite{Farzanehfar2021} subsequently reproduced this finding to country-scale datasets, finding that four spatio-temporal points still uniquely identified up to 93\% of individuals. This uniqueness is not necessarily confined to fine-grained GPS traces: Zang and Bolot~\cite{Zang2011} showed 
that even coarse-grained location data based on mobile call records retains enough uniqueness to enable re-identification. Spatial and temporal patterns in location data can reveal points of interest (POIs), such as home, workplace, or leisure venues, creating substantial opportunities for re-identification and user profiling even when coordinates are imprecise \cite{Primault2019,wiedemann2024you}. By using information associated with POIs, such as exact addresses, an adversary can conduct web searches to infer identifying information, including the name of a person registered at a specific residential address~\cite{krumm2007inference}.

A methodological precedent close to ours is Krumm's 2007 study~\cite{krumm2007inference}, which extracted home locations from two-week GPS traces of 172 drivers, ran the inferred coordinates through a free white-pages reverse lookup, and successfully named about 5\% of subjects. Krumm explicitly identified two technological bottlenecks: poor reverse geocoding (only 1.2\% of pings matched the true home address) and incomplete white-pages coverage (43\% of self-reported addresses returned no listing). Our work extends this line of research by replacing the four hand-coded inference heuristics and the single reverse-lookup call with an LLM-agent pipeline that exploits the breadth of modern web.

\paragraph{AI agents and automated OSINT.}
Re-identifying individuals at scale from spatio-temporal mobility traces has 
historically required sophisticated technical knowledge, including API configuration, data preprocessing, and familiarity with OSINT techniques. We argue that, in the age of AI, these barriers are significantly lowered, expanding the pool of potential adversaries who could misuse these capabilities. Staab et~al.~\cite{Staab2024Beyond} showed that LLMs can infer personal attributes such as location, income, and sex from short pieces of public text, achieving up to 85\% top-1 accuracy while operating at roughly $100\times$ lower cost and $240\times$ faster than human investigators -- a first concrete quantification of the speed and cost advantage that makes large-scale privacy attacks practical. Building on this attribute-inference paradigm, Lermen et~al.~\cite{Lermen2026} demonstrated that LLM agents can already autonomously de-anonymise users at scale. Their agents summarized anonymised Hacker News profiles, searched the web, cross-referenced sources, and reasoned over the evidence to infer likely real-world identities. Although their attack targets text-based online identities rather than GPS microdata, the underlying paradigm shift is the same: a task previously requiring hours of manual expert effort becomes scalable and inexpensive through agentic tool use.

Recent technological advances further show that location privacy is not limited to explicit GPS data; sensitive location information can also be inferred from images. Luo et~al.~\cite{Luo2026Doxing} showed that adversaries can use multimodal large reasoning models to infer geolocation information from user-generated images, such as selfies shared on social media platforms. Such inferences may expose sensitive places, including a user’s home address or frequently visited leisure locations.
\section{Methodology}
\label{sec:methodology}

\subsection{Agentic Re-Identification Pipeline}
\label{subsec:pipeline}

\paragraph{Threat model.}
We adopt a threat model in which an adversary has obtained raw GPS mobility trace data for a single device -- for instance, by purchasing it from a commercial data broker -- and seeks to recover the real-world identity of the device's primary user. In the disclosure-control taxonomy this corresponds to \emph{identity disclosure}~\cite{Duncan_1989_Risk}. 
The adversary has access only to freely available data on the public web (e.g., search engines, online directories, public registers, social media, news, and business registers) and to LLM agents. No fake social media profiles, social engineering, or any other form of deception is used to gain access to additional information. A re-identification is judged successful only when the trace is linked to a specific named individual. Thus, the harm extends beyond merely naming the device-holder. Once a candidate is named, every subsequent ping in the trace ceases to be an anonymised data point and becomes a movement diary attached to that person -- for example, a local politician’s weekly visits to the ice-hockey rink, the Friday afternoon drop-off at a children’s football club, or the Monday-evening table at a named restaurant.

What is novel in our setting is not this threat model but its operationalisation: prior work assumed human-analyst effort as the practical bottleneck, whereas we equip the adversary with autonomous LLM agents that can execute the same attack at scale and at orders-of-magnitude lower cost than a human investigator~\cite{Staab2024Beyond,Lermen2026}.

\begin{figure}[p]
  \centering
  \includegraphics[height=1\textheight]{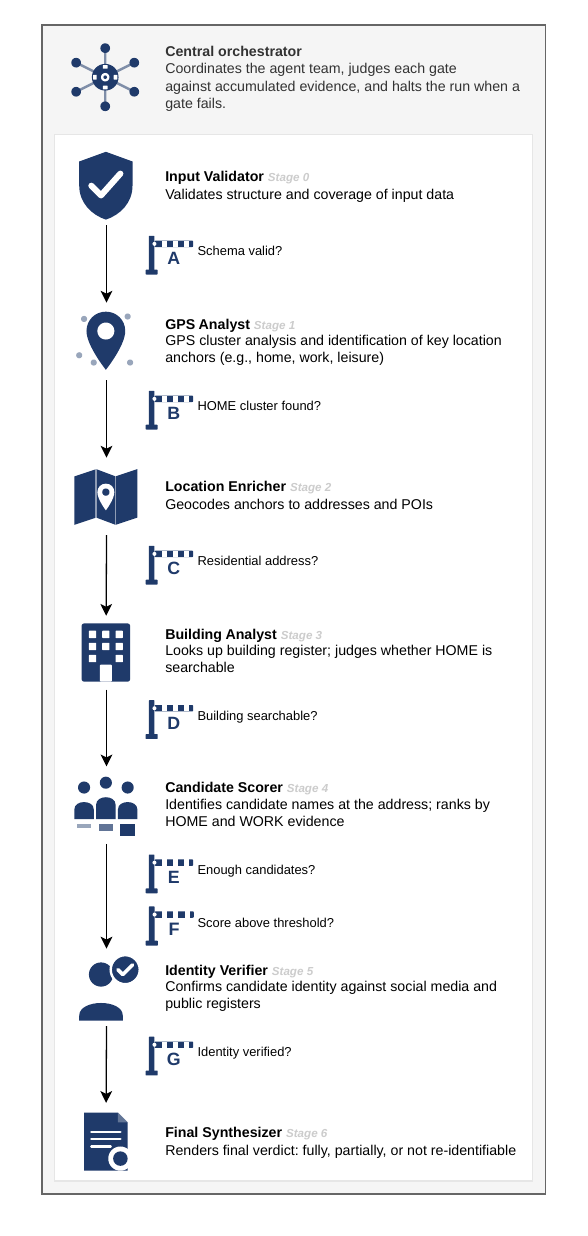}
  \caption{The agentic re-identification pipeline: seven specialist agents (icons) connected by quality gates (barriers). The central orchestrator routes stage outputs, enforces gates, and maintains a running uncertainty ledger; the run halts whenever a gate's criteria are not met. Agent and gate icons designed with the assistance of Claude Opus 4.7 (Anthropic, 2026).}
  \label{fig:pipeline}
\end{figure}

The pipeline decomposes the re-identification task into seven specialist stages controlled by a central orchestrator, grouped into three conceptual phases (\Cref{fig:pipeline}). Each stage produces structured output consumed by the next; the orchestrator enforces quality gates between transitions and halts the run early when evidence falls below a sufficiency threshold.

\begin{figure}[H]
  \centering
  \includegraphics[width=1\linewidth]{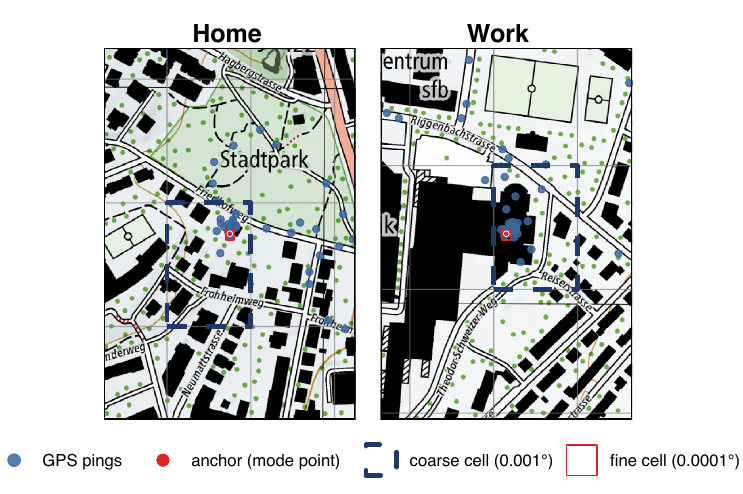}
  \caption{Stage~1 grid-based mode finding on a simulated fictitious trace, illustrating the home cluster (left) and the work cluster (right). Grey grid lines mark the $0.001^\circ \times 0.001^\circ$ cells the GPS-analyst agent rounds pings into ($\approx 111\,\text{m} \times 76\,\text{m}$). The coarse cell with the most pings is highlighted (dashed blue); within it, the agent re-bins pings on a finer $0.0001^\circ \times 0.0001^\circ$ grid ($\approx 11\,\text{m} \times 7.6\,\text{m}$) and picks the densest fine cell (solid red) as the cluster's mode. A representative coordinate inside that cell (red dot) is the anchor passed to the downstream stages. Background: \copyright\ Federal Office of Topography swisstopo.}
  \label{fig:poi_clusters}
\end{figure}

In the \emph{spatial-analysis} phase, Stage~0 validates the input and Stage~1 (\emph{GPS Analyst}) extracts HOME and WORK anchors using grid-based mode-finding on the raw trace (see \Cref{fig:poi_clusters}), classifying them by temporal pattern (night/weekend-dominant locations indicate likely HOME, while weekday daytime-dominant locations indicate likely WORK). The \emph{location-attribution} phase (Stages~2--3) grounds these anchors in real-world addresses and assesses their residential specificity: Stage~2 (\emph{Location Enricher}) reverse-geocodes each anchor against several derived coordinates, producing an address candidate with consensus confidence and a bundle of plausible alternatives when geocodes disagree. For WORK clusters it additionally enumerates named institutions near the mode point as candidate employers. Stage~3 (\emph{Building Analyst}) queries the Swiss Federal Register of Buildings and Dwellings (GWR)~\cite{SwissGWR} for the HOME building and sets a ceiling on the attribution strength that downstream stages may reach. A single-family dwelling permits \textit{full re-identification}, a small multi-unit building caps the result at \textit{partial re-identification} unless later evidence is unusually strong, and large apartment blocks or non-residential structures may halt the pipeline with a \textit{not re-identifiable} verdict. In the \emph{identity-attribution} phase (Stages~4--6), candidate identities are identified and validated. Stage~4 (\emph{Candidate Scorer}) ranks individual candidates from public directories and registers using both HOME and WORK evidence; Stage~5 (\emph{Identity Verifier}) seeks independent corroboration for the top candidate in social-media and public sources as a second-opinion check. Finally, in Stage~6, the \emph{Final Synthesizer} aggregates the accumulated evidence and renders the final attribution outcome.

A central design principle is strict monotonicity: earlier stages establish geometric and address-level anchors that later stages may enrich but not revise. This ensures that the attribution chain remains traceable and prevents circular reasoning between the spatial and identity phases.

Explicit quality gates between stages prevent weak or ambiguous evidence from propagating forward. If the address evidence fails to meet a residential-specificity threshold, the pipeline halts before identity search begins. Likewise, identity candidates that cannot be positively corroborated by behavioural evidence are not passed to the synthesis stage. A running uncertainty ledger tracks unresolved ambiguities from each stage and is surfaced in the final output, making the limits of each inference explicit.

The pipeline thus follows a conservative, evidence-chained attribution logic: each phase can only proceed on the basis of positive evidence from the prior phase, and the final output reflects not only the strongest achievable attribution but also the conditions under which that attribution fails.

\subsection{Study Design and Simulated Data}
\label{subsec:study_design}

This work sits in inherent tension between methodological rigour and individual privacy: empirically establishing the attack in a real-world setting requires performing it on real individuals -- potentially compromising their privacy.

We address this tension as follows: First, the sample is consent-based, and location information is included only for participants who gave explicit consent. Second, GPS traces are simulated around participants’ real home and work addresses, rather than obtained from real data brokers. Third, re-identification relies only on publicly available information, without the use of fake profiles, social engineering, or any form of deception. Finally, we do not release the simulated dataset, the underlying code, the prompts, or the agent skills.

To evaluate re-identification under controlled and ethically grounded conditions, we constructed a simulated GPS dataset derived from the real home and work addresses of study participants. All participants explicitly consented for their home and work address data to be used in re-identification experiments. No pre-screening for online visibility or public linkability was applied, so the sample includes individuals across the full spectrum of digital footprint -- from those who appear prominently in public registers and employer websites to those with no retrievable online presence.

Home and work locations constitute the two most discriminative spatial anchors for re-identification: prior work has shown that only a handful of spatio-temporal points suffice to uniquely identify the vast majority of individuals in a mobility dataset, with home and work being the most informative pair~\cite{Golle2009,Stroebl2025}. We therefore simulated location clusters around participants’ true home addresses and, where applicable, their true work addresses, with GPS pings generated around each anchor using spatial jitter to reflect realistic device noise; \Cref{fig:poi_clusters} illustrates one such pair of clusters.

Traces are simulated over a three-week period. Pings cluster at the home anchor during morning and evening hours and at the work anchor during the day; leisure stops are short-duration clusters positioned near but distinct from the two anchors, suppressed at night so that night-time pings remain predominantly at home. Leisure stops do not correspond to participants’ actual locations. Instead, they are introduced as noise to increase difficulty to infer participants’ home and workplace locations. Each trace also incorporates random dropouts, burst-sampling artefacts at stable locations, and a small fraction of out-of-order timestamps, mimicking real-world data quality. The simulation parameters (i.e., sampling cadence, spatial accuracy distribution, home/work anchoring patterns, and dropout characteristics) are informed by a real commercial broker dataset that the authors examined for this purpose, so that the simulated traces qualitatively resemble what an adversary would actually receive. No individuals in the broker dataset were re-identified, and the dataset is not used in any other part of the study.

\paragraph{Sample description.}
A convenience sample of $n = 43$ participants was recruited in Switzerland. Of these, 21 participants lived in single-household buildings and 22 lived in multi-household buildings. Publicly available online information varied across participants: 19 were listed in a public residential register, 17 had their name and home address listed on other websites, 6 maintained a personal website, 32 had a LinkedIn profile, and 33 were listed by name on a company website.

\paragraph{Ground truth and scoring.}
Ground truth consisted of each participant's self-reported real name, home address, and work address, where applicable, collected at the time of consent, together with their online presence as used to assess re-identifiability. Each pipeline run produces one of three result types: \textit{full re-identification}, \textit{partial re-identification}, or \textit{not re-identifiable}. A \textit{full re-identification} verdict was assigned when the spatio-temporal data could be unambiguously linked to a named candidate. A \textit{partial re-identification} verdict was assigned when the pipeline narrowed the candidate set to a small group of co-residents at the correct address and household, but could not identify a single individual. A \textit{not re-identifiable} verdict was assigned when the available evidence was insufficient to support either full or partial re-identification. For each participant, the authors manually verified the degree of re-identifiability through web search against which pipeline performance is reported in \Cref{sec:results}.

\paragraph{Pipeline development and implementation.}
The pipeline was developed against a separate development set of six GPS test traces: agent prompts, quality-gate thresholds, and orchestrator stop rules were refined iteratively by inspecting per-stage outputs and adjusting whenever a stage produced an incorrect attribution or an unjustified escalation. The 43 participant traces were used only for the final evaluation reported in \Cref{sec:results}; no parameter, prompt, or threshold was modified in response to evaluation-set outcomes.

The pipeline is orchestrated by Anthropic's Claude Code command-line interface (CLI)~\cite{ClaudeCode2026} in non-interactive mode (\texttt{claude -p}), running the orchestrator and stage specifications of \Cref{subsec:pipeline} within a single tool-using session. Runs use \texttt{claude-sonnet-4-6} as the primary model, with the CLI's default settings for temperature, top-p, and other generation controls; pre-authorised tools include web search, web fetch, and local file read/write. Per-run JSON and CSV logs capture stage outputs, token counts, and wall-clock time, from which the cost and runtime figures in \Cref{sec:results} are computed directly. A bash script parallelises one session per device CSV file at a concurrency of four on a Linux server. To prevent information leakage, all ground-truth participant information was stored in a separate, isolated environment inaccessible to the agents.
\section{Results}
\label{sec:results}
We applied the pipeline to the 43 participant traces described in \Cref{subsec:study_design}, scoring each run against the manually established re-identifiability label. 
Table \ref{tab:confusion} summarises the results of these re-identification attempts.
Full re-identification was achieved in 18 of 43 cases (41.9\%). Among 
the 19 cases in which the pipeline returned a named candidate, 18 matched the 
ground truth, corresponding to a full re-identification precision of 94.7\%. 
Conversely, among the 25 cases in which full re-identification was possible, the 
pipeline recovered the name of the individual in 18 cases (72.0\%); the remaining
seven cases were wrongly classified as partially ($n = 2$) or not re-identifiable ($n = 5$). 

However, in the two cases, in which the agentic re-identification attempt achieved only partial rather than full re-identification, the predicted household link included the correct household, enabling the adversary to infer at least the individual's correct last name. Furthermore, in two of the five cases in which full re-identification was possible but the pipeline returned no re-identification, the correct candidate name had been found. However, the pipeline concluded that the evidence was insufficient, either because workplace verification was unavailable or because the inferred home address corresponded to a multi-unit building.

For the 16 ground-truth cases that were not re-identifiable, the pipeline correctly halted without naming a candidate in 14 instances (87.5\%). One case nonetheless produced a named-candidate output owing to an incorrect match between a candidate name and home address, and one produced partial re-identification in which the identified household was incorrect because the house number had been wrongly inferred from the GPS data. Regarding re-identifiability, participants with a higher online footprint -- in particular those listed in a public residential register -- were more readily re-identified than those with limited online presence.

\begin{table}[!htbp]
  \centering
  \caption{Confusion matrix: ground-truth outcome (rows) vs.\ re-identification pipeline outcome (columns); $n = 43$ evaluation cases.}
  \label{tab:confusion}
  \setlength{\tabcolsep}{6pt}
  \begin{tabular}{@{}l ccc c@{}}
    \toprule
    & \multicolumn{3}{c}{\textbf{Re-Identification Pipeline Outcome}} & \\
    \cmidrule(lr){2-4}
    & \shortstack{Full\\re-identification} & \shortstack{Partial\\re-identification} & \shortstack{Not\\re-identifiable} & \textbf{Total} \\
    \midrule
      Full re-identification      & 18 &  2 &  5 & 25 \\
      Partial re-identification   &  0 &  2 &  0 &  2 \\
      Not re-identifiable         &  1 &  1 & 14 & 16 \\
    \midrule
      \textbf{Total}              & 19 &  5 & 19 & 43 \\
    \bottomrule
  \end{tabular}
  \par\smallskip
  \begin{minipage}{\linewidth}
    \footnotesize
    \textit{Full re-identification}: pipeline returned a uniquely identified individual;
    \textit{Partial re-identification}: household confirmed, but shared by multiple residents -- no individual uniquely identifiable;
    \textit{Not re-identifiable}: pipeline halted without naming a candidate.
  \end{minipage}
\end{table}

\enlargethispage{2\baselineskip}
\paragraph{Cost and run-time.}
Across the 43 evaluation runs, each attempt cost on average \$2.24 in API charges at list prices, produced on average 58.5K output tokens (range: 39.5K--78.9K), and took on average 17 minutes of unattended computation (range: 11.7--28.5 minutes). The pipeline parallelises freely; all 43 runs completed in a single day at a concurrency of four. No controlled human-analyst baseline was conducted.
\section{Discussion and Conclusion}
\label{sec:conclusion}

The agentic re-identification pipeline presented in this study successfully re-identified 72\% of cases for which unambiguously re-identification was possible. For comparison, Krumm's 2007 inference attack succeeded on roughly 5\% of subjects, limited by the technological maturity of geocoders and directories at the time~\cite{krumm2007inference}. At an average of 17 minutes and \$2.24 per attempt (\Cref{sec:results}), and with free parallelism, the marginal cost of mass re-identification is low realtive to manual human effort. A manual attempt by a human analyst would demand continuous attention and would likely take longer -- plausibly considerably longer without specialised tooling -- though we did not measure human performance directly. The decisive advantage of the agentic pipeline is therefore not raw per-case speed but the combination of low cost, wall-clock parallelism, and the elimination of human involvement, consistent with the cost advantage Staab et~al.~\cite{Staab2024Beyond} report for LLM-driven inference. 

Although building such a pipeline still requires some technical expertise, these barriers are substantially reduced as AI-assisted software development lowers the skill threshold for implementing, adapting, and scaling similar workflows~\cite{Fawzy2026VibeCoding}. As these technologies continue to improve, constructing and operating systems to re-identify individuals will become increasingly accessible over time, posing new challenges for privacy. 

We call on the SDC community to develop adversarial risk models that explicitly account for the capabilities and scale of agentic AI attacks. Defenders should adopt SDC techniques that come with formal guarantees rather than relying on de facto anonymity claims. Although GDPR Recital~26~\cite{GDPR2016Recital26} and subsequent CJEU jurisprudence~\cite{CJEUIAB2024} classify pseudonymised data including location traces as personal data when re-identification is reasonably likely, broker data is routinely released under \emph{anonymisation} claims that conflate the two~\cite{FTCGravy2024,FTCMobilewalla2024,MeineckDachwitz2024}.

\paragraph{Limitations.}
For ethical reasons, and consistent with prior work~\cite{Li2026Anthropic,Lermen2026}, no exact prompts or code used in this study is disclosed, as doing so could facilitate misuse, including the targeting of real individuals. The simulated dataset is similarly withheld: although it is anchored on consenting participants, the per-trace spatio-temporal resolution is high enough that release would carry residual re-identification risk for the participants themselves.

Three bounds apply to the generalisability of our findings. The simulated traces share a single generator that is informed by -- but not statistically calibrated against -- a real broker dataset; the evaluation is scoped to Switzerland and to country-specific directory and register sources; and the sample is small relative to broker datasets containing millions of devices, making the reported re-identification rates a proof-of-concept feasibility result rather than a population estimate. The country-specificity bound is particularly relevant for Stage~3, which relies on the Swiss Federal Register of Buildings and Dwellings (GWR)~\cite{SwissGWR} for per-building category and dwelling counts. Equivalent national registers exist in many other countries and would support the same gate logic, but where they do not the pipeline would have to fall back to less precise proxy data.

Compared with real data broker data, our simulation anchors only home and work locations. Real broker traces may also include leisure-location dwell patterns, which an adversary could cross-reference with public membership records or event listings, thereby potentially increasing the risk of re-identification further. However, real data may also contain noisier GPS points, making it more difficult to pinpoint exact addresses.

Several methodological caveats further bound our conclusions. We did not run a non-agentic baseline, so the cost and time advantages discussed above are anchored to Staab's text-domain measurements \cite{Staab2024Beyond} and to our own informal observation rather than to a controlled comparison. Frontier LLMs include safety filters that constrain privacy-sensitive operations; our results are conditional on the specific safety posture of the model versions used~\cite{Li2026Anthropic,Lermen2026,Ko2026}. Because each evaluation case was run only once, we did not quantify the run-to-run variability of pipeline outcomes. Finally, the public directory sites our pipeline queries impose access controls and anti-crawling measures whose impact at higher scale our small-sample evaluation does not capture.

\paragraph{Future Work and the Escalating Threat Horizon.}
The re-identification pipeline presented in this paper was limited to searching and reading publicly available information. A technically feasible and more concerning next step would be for an adversary to extend the agent’s capabilities to collect non-public or semi-public information. This could include the use of more sophisticated tooling, such as image geolocation \cite{Luo2026Doxing}, people-search services, and facial recognition~\cite{Acquisti2014Face,Wang2025Protego}. It could also involve autonomous management of fake social-media accounts, direct outreach to individuals or companies via e-mail or even telephone, or other forms of social engineering~\cite{Kumarage2025SE,Figueiredo2025ViKing}. Such capabilities could substantially improve the adversary’s ability to identify and verify candidates.

A second dimension concerns local execution: open-weight LLMs in the 30--65B parameter range can plausibly carry the structured stages on commodity hardware~\cite{Grattafiori2024Llama3,Dettmers2023QLoRA}. Local execution could be used to reduce the per-target cost floor and remove the API audit trail, undermining detection strategies that rely on hosted-model logs or rate-limit signatures. Taken together, these trends point toward a threat horizon in which such attacks become progressively cheaper and more capable. The statistical disclosure control community should therefore treat agentic AI adversaries as a real rather than a hypothetical threat.

\paragraph{Acknowledgements.}
This work was funded by the Swiss National Science Foundation with grant 
\textit{"Harnessing event and longitudinal data in industry and health sector 
through privacy preserving technologies"} (grant number 211751).

\paragraph*{Disclosure of Interests.} 
The authors have no competing interests to declare that are relevant to the 
content of this article.

\bibliographystyle{splncs04}
\bibliography{bibliography/references}


\end{document}